%% file: main.tex
\def\isarxiv{1} %%%Arxiv version, we uncomment this line

\ifdefined\isarxiv
\documentclass[11pt]{article}
\else
 
\documentclass[conference]{IEEEtran}
\IEEEoverridecommandlockouts
\usepackage{hyperref}

\def\BibTeX{{\rm B\kern-.05em{\sc i\kern-.025em b}\kern-.08em
    T\kern-.1667em\lower.7ex\hbox{E}\kern-.125emX}}
\fi

\usepackage[utf8]{inputenc}

\usepackage{microtype}
\usepackage{amsmath}
\usepackage{amsthm}
\allowdisplaybreaks
\usepackage{amssymb}
\usepackage{algorithm}
\usepackage{subfig}
\usepackage{color}
\usepackage{epstopdf}
\usepackage{url}
\usepackage{graphicx}
\usepackage{color}
\usepackage{epstopdf}
\usepackage{algpseudocode}
\usepackage{scrextend}
\usepackage{bbm}
\usepackage{comment}

\usepackage{multirow}
\usepackage{dsfont}
\usepackage{mathtools}
\usepackage{enumitem}

\usepackage[makeroom]{cancel}
\usepackage{stmaryrd}
\usepackage{booktabs, makecell}
\usepackage{pifont}% http://ctan.org/pkg/pifont
\usepackage{lipsum}

\usepackage{amsfonts}       % blackboard math symbols
\usepackage{nicefrac}       % compact symbols for 1/2, etc.
\usepackage{xcolor}         % colors

\usepackage{tikz}
\usepackage{hyperref}
\definecolor{mydarkblue}{rgb}{0,0.08,0.45}
\hypersetup{colorlinks=true, citecolor=mydarkblue,linkcolor=mydarkblue}

\usetikzlibrary{arrows}
\ifdefined\isarxiv
\usepackage[margin=1in]{geometry}
\else

\fi
\graphicspath{{./figs/}}

\definecolor{b2}{RGB}{51,153,255}
\definecolor{mygreen}{RGB}{80,180,0}
\definecolor{yl}{RGB}{255,80,0}
\definecolor{myl}{RGB}{180,80,20}

\theoremstyle{plain}
\newtheorem{theorem}{Theorem}[section]
\newtheorem{lemma}[theorem]{Lemma}

\newtheorem{assumption}[theorem]{Assumption}

\theoremstyle{definition}
\newtheorem{definition}[theorem]{Definition}
\newtheorem{example}[theorem]{Example}

\theoremstyle{remark}
\newcommand{\wh}{\widehat}
\newcommand{\wt}{\widetilde}

\newcommand{\E}{\mathbb{E}}

\newcommand{\R}{\mathbb{R}}

\renewcommand{\hat}{\wh}

\newcommand{\tr}{\mathrm{tr}}

\newcommand{\Tmat}{{\cal T}_\mathrm{mat}}

\newcommand{\supp}{\mathrm{supp}}
\newcommand{\append}{\mathrm{append}}

\ifdefined\isarxiv
\title{Sparse Convolution for Approximate Sparse Instance}

\date{}

\author{
Xiaoxiao Li\thanks{
\texttt{xiaoxiao.li@ece.ubc.ca}. University of British Columbia.}
\and 
Zhao Song\thanks{\texttt{zsong@adobe.com}. Adobe Research.}
\and 
Guangyi Zhang\thanks{
\texttt{guangyi.zhang@mail.mcgill.ca}. McGill University.}
}

\else 

\fi

\begin{document}

%\ifdefined\isicml

\ifdefined\isarxiv

\begin{titlepage}
\maketitle
\begin{abstract}
\input{abstract}

\end{abstract}
\thispagestyle{empty}
\end{titlepage}

\else

\maketitle
Nothing

\fi
\input{intro}
\input{related_work}

\input{preli}
\input{sparse_conv}
\input{iterative_correct_error}

\input{conclusion} 

\ifdefined\isarxiv 
\bibliographystyle{alpha}
\else 
\bibliographystyle{IEEEtran}
\fi
\bibliography{ref}

\end{document}

%% file: abstract.tex
Computing the convolution $A \star B$ of two vectors of dimension $n$ is one of the most important computational primitives in many fields. For the non-negative convolution scenario, the classical solution is to leverage the Fast Fourier Transform whose time complexity is $O(n \log n)$. However, the vectors $A$ and $B$ could be very sparse and we can exploit such property to accelerate the computation to obtain the result. In this paper, we show that when $\|A \star B\|_{\geq c_1} = k$ and $\|A \star B\|_{\leq c_2} = n-k$ holds, we can approximately recover the all index in $\supp_{\geq c_1}(A \star B)$ with point-wise error of $o(1)$ in $O(k \log (n) \log(k)\log(k/\delta))$ time. We further show that we can iteratively correct the error and recover all index in $\mathrm{supp}_{\geq c_1}(A \star B)$ correctly in $O(k \log(n) \log^2(k) (\log(1/\delta) + \log\log(k)))$ time.

%% file: intro.tex
\section{Introduction}

Computing the convolution $A \star B$ of two vectors of dimension $n$ is one of the most important computational primitives. 
It also has been widely used in many fields such as computer vision~\cite{lbs+18,czp+18,tl19,xsm+21}, signal processing~\cite{k81,sbhl98}, graph mining~\cite{cwh+20}.
For example, it has applications in problems like three number sum problem and all-pairs shortest paths, where the entries of vector $A$ and $B$ are non-negative integers. 
In string algorithms, non-negative convolution is employed when computing the Hamming distance of a pattern as well as the distance between each sliding window of a text~\cite{fp74}. 
%%%%%%%%%%%%%%%%%
The classical algorithm to compute the non-negative convolution leverages the Fast Fourier Transform (FFT) and its running time complexity is $O(n \log n)$.
Algorithms in \cite{cl15} give $O(n^2)$ for 3-SUM and related problems. 
Thanks to the key techniques mentioned in \cite{cl15}, the celebrated Balog-Szemeredi-Gowers Theorem (BSG Theorem) \cite{bal07,ssv15} and FFT, the authors in \cite{cl15} gave the first truly subquadratic algorithms for various problems associated with 3SUM.

The key techniques come from BSG Theorem \cite{bal07,ssv15} and FFT. The BSG theorem has been improved by \cite{s15}, however the result of \cite{s15} did not provide efficient algorithm (with small running time) for solution. They only show the existence of such solution.

However, in many scenarios, the vector $A$ and $B$ could be very sparse and we can exploit such sparsity to improve the running time complexity compared to the classical algorithm implemented via FFT algorithm. \cite{bfn21} studied the exact $k$-sparse case and proves that $k$-sparse non-negative convolution can be reduced to dense non-negative convolution with an additive $k \log \log k$ term. 
We consider the problem of approximately $k$-sparse non-negative convolution and state the assumption as follows: 
\begin{assumption}[Approximate sparse non-negative convolution]\label{ass:approx_sparse}
Assume $A,B \in \R_+^n$. Additionally, there exist $c_1 = \Omega(1)$ and $c_2 = o(n^{-2})$ such that $\|A \star B\|_{\geq c_1} = k$ and $\|A \star B\|_{\leq c_2} = n-k$.
\end{assumption}
Note that the previous work~\cite{bfn21} only considers $c_2= 0$. We can handle some error here.

We summarize our contributions as follows:
\begin{itemize}
    \item We study the approximately $k$-sparse non-negative convolution problem and design an approximate sparse convolution algorithm (Algorithm~\ref{alg:approx_sparse_conv}) such that it can approximately recover the all index in $\supp_{\geq c_1}(A \star B)$ with point-wise error of $o(1)$ in 
    $O(k \log (n) \log(k)\log(k/\delta))$ time.
    \item We further design another algorithm (Algorithm~\ref{alg:iterative_correct}) which can iteratively correct the error and recover all index in $\supp_{\geq c_1}(A \star B)$ correctly in \\$O(k \log(n) \log^2(k) (\log(1/\delta) + \log\log(k)))$ time.
\end{itemize}

\textbf{Roadmap} We first discuss the related work in Section~\ref{sec:related_work}. We then present some preliminary definitions and lemmas in Section~\ref{sec:prelim}. We present our result for approximate sparse convolution in Section~\ref{sec:approx_sparse_conv}. We then show how to iteratively correct the error in Section~\ref{sec:approx_sparse_conv_iter_correcting_error}. We conclude our paper in Section~\ref{sec:conclusion}.

%% file: related_work.tex
\section{Related Work}\label{sec:related_work}

\paragraph{Sparse Convolution} There has been a lot of previous work on accelerating sparse convolution computation~\cite{m95,ch02,r08,mp09,r18,n20}. 

Hash function has wide applications.
In sparse convolution problem: for vectors $u, v \in \R_+^n$, compute their classical convolution $\vec{u} * \vec{v}=\vec{z}$ (where $z_k=\sum_{i=0}^k u_i v_{k-i}$ ) in "output-sensitive" time, close to $\|\vec{z}\|$, the number of nonzeros in $\vec{z}$. The problem was raised by Muthukrishnan \cite{m95}, and previously solved by Cole and Hariharan \cite{ch02}.

Cole et. al. \cite{ch02} obtained a $O(k \log^2(n) + \mathrm{polylog}(n))$ time complexity for sparse non-negative convolution case with a Las Vegas algorithm. Their strategy incorporates a number of concepts, including encoding characters with complex entries before using convolution, and constructs on linear hashing and string algorithms to identify $\supp(A\star B)$.
Recent methods \cite{r08,n20,r11,vjl12,vjl13,ar15,ggc20} rely largely on hashing modulo an arbitrary prime number.

This method loses one log factor as a result of the Prime Number Theorem's stipulation for the density of primes and obtained $O(k\log k)$ or even $O(k\log^2 k)$. Nakos et.al.\cite{n20} achieves $\wt{O}(k \log^2(n) + \mathrm{polylog}(n))$ time complexity for sparse general convolution case. There are several implementation for sparse convolution algorithms in~\cite{mp14,mp15}.
Sparse convolution is also related to sparse Fourier transform, which has been also widely studied~\cite{ggims02,hikp12}.

\paragraph{Sparse Matrix Multiplication} 
If most elements in a matrix are zero, we call this matrix is a sparse matrix. However, it would be a waste of space and time to save and compute these sparse matrices. Therefore, it is important to only save the nonzero elements. Standardly in sequential programming languages, sparse matrices are represented using an array with one element per row, each of which comprises a linked list of the nonzero values in that row and their column number. 
Sparse matrix multiplication is a compute kernel used in a variety of application fields. It plays an important role in many areas such as data analytics, graph processing, and scientific computing.
In the past decades, there has been many previous work on algorithm side optimization~\cite{yz05,bg12,wls18} and hardware acceleration~\cite{lv14,zwhd20,sln15}. The appearance of these algorithms accelerates the sparse matrices manipulation greatly.

%% file: preli.tex
\section{Preliminary}\label{sec:prelim}

\paragraph{Notations} For any natural number $n$, we use $[n]$ to denote the set $\{1,2,\dots,n\}$.
We use $A^\top$ to denote the transpose of matrix $A$.
For a probabilistic event $f(x)$, we define ${\bf 1}\{f(x)\}$ such that ${\bf 1}\{f(x)\} = 1$ if $f(x)$ holds and ${\bf 1}\{f(x)\} = 0$ otherwise.

We use $\Pr[\cdot]$ to denote the probability, and use $\E[\cdot]$ to denote the expectation if it exists. 
For a matrix $A$, we use $\tr[A]$ for trace of $A$. We use $\Tmat(a,b,c)$ to denote the time of multiplying an $a \times b$ matrix with another $b \times c$ matrix.

We then provide several definitions of $\partial$ notation, non-negative convolution, cyclic convolution, generalized norm and support, and rounding here.

\begin{definition}[The $\partial$ notation]
Given a vector $A\in \R^n$, we define $\partial A$ to be the vector of the same dimension such that 
\begin{align}\label{eq:def_partial}
    (\partial A)_i := A_i \cdot i.
\end{align}
\end{definition}

\begin{example}
We can choose $A$ to be a length $7$ vector, $A =(3,1,2,1,2,1,1)$. Then we can compute $\partial A$, which becomes $\partial A =(3,2,6,4,10,6,7)$.
A visualization  of $A$ and $\partial A$ is shown in Figure~\ref{fig:AdA}.
\begin{figure}[H]
    \centering
    \includegraphics[width=0.45\textwidth]{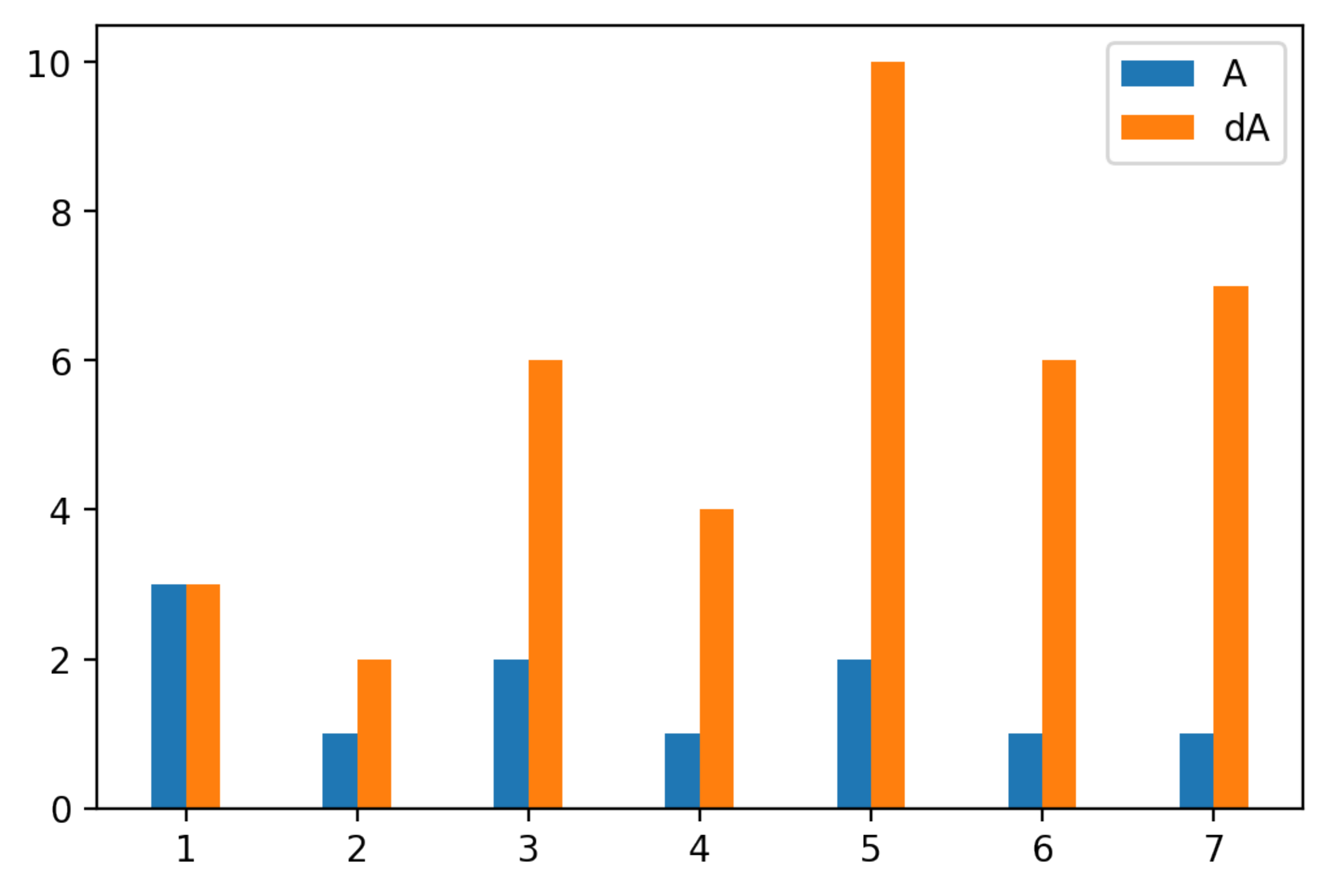}
    \caption{Visualization of $A$ and $\partial A$}
    \label{fig:AdA}
\end{figure}
\end{example}

\begin{definition}[Non-negative Convolution]
Given vectors $A,B \in \mathbb{N}^n$, the vector $C = A \star B \in \mathbb{N}^{2n-1}$ is defined by 
\begin{align*}
    C_k := \sum_{i=0}^{n} A_i \cdot B_{k-i}.
\end{align*}

\end{definition}

\begin{figure*}[!ht]
\subfloat[Example~\ref{ex:ABC_1}]{\includegraphics[width=0.32\textwidth]{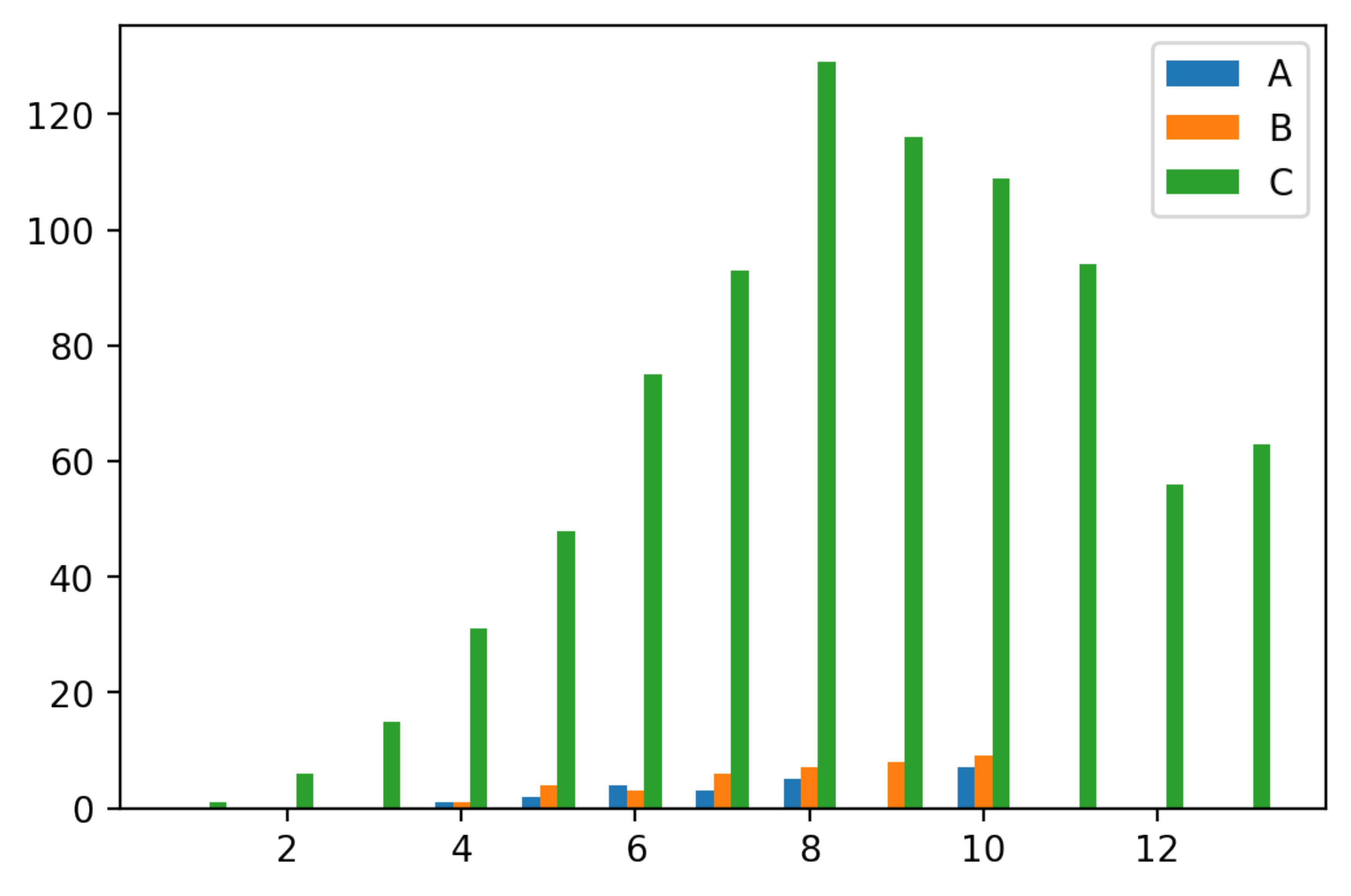}\label{fig:ABC}} 
\subfloat[Example~\ref{ex:ABC_2}]{\includegraphics[width=0.32\textwidth]{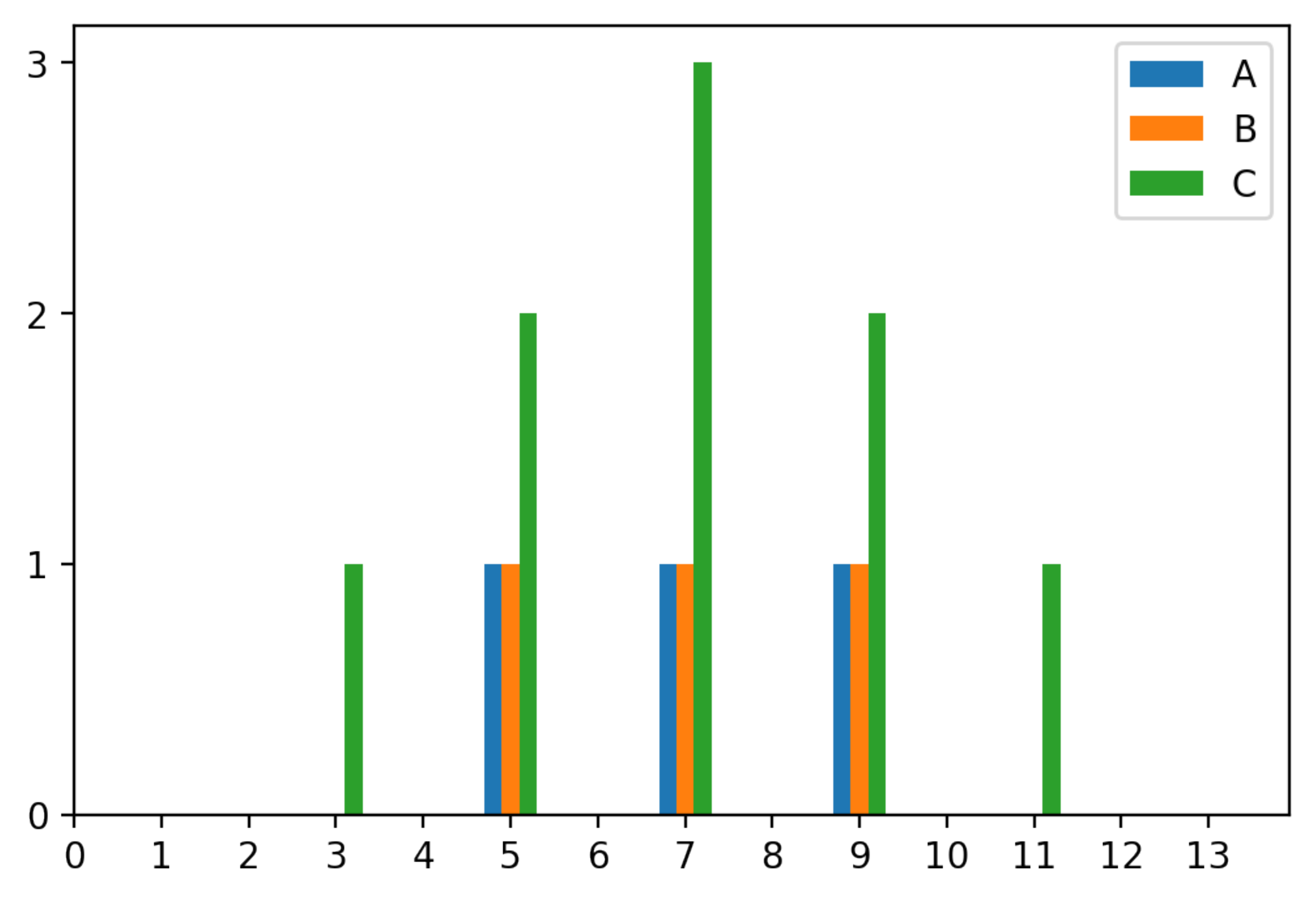}\label{fig:ABC_2}}
\subfloat[Example~\ref{ex:ABC_3}]{\includegraphics[width=0.32\textwidth]{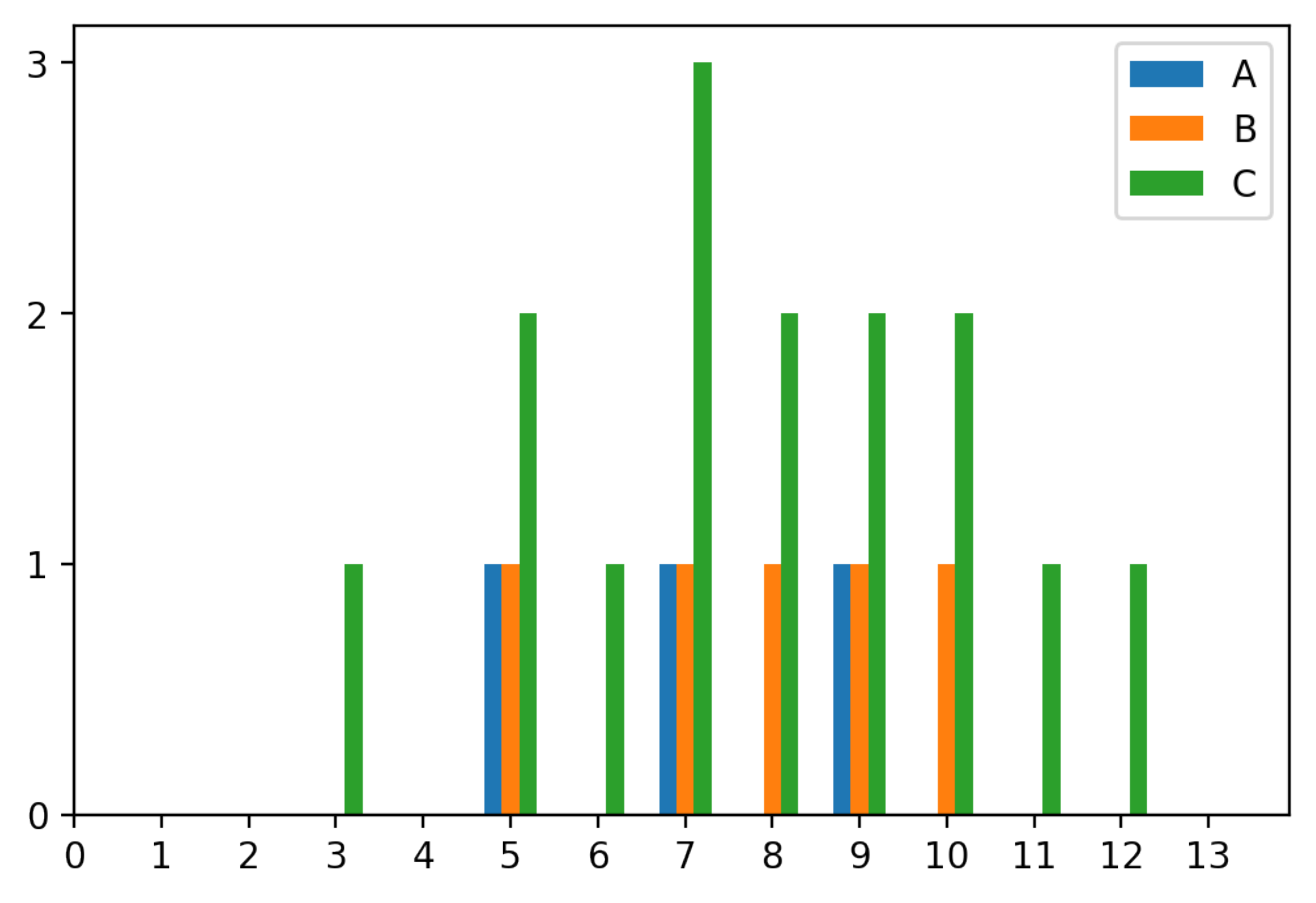}\label{fig:ABC_3}} 
\caption{Visualization of $A$, $B$ and $C$.}
\label{fig:ABC_All}
\end{figure*}

\begin{example}\label{ex:ABC_1}
If $A = (1,2,4,3,5,0,7)$, $B = (1,4,3,6,7,8,9)$, then\\ $C=(1, 6, 15, 31, 48, 75, 93, 129, 116, 109, 94, 56, 63)$.
A visualization  of $A$, $B$ and $C$ is shown in Figure~\ref{fig:ABC}.
\end{example}
\begin{example}\label{ex:ABC_2}
If $A = (0,1,0,1,0,1,0)$, $B = (0,1,0,1,0,1,0)$, then $C=(0, 0, 1, 0, 2, 0, 3, 0, 2, 0, 1, 0, 0)$.
A visualization of $A$, $B$ and $C$ is shown in Figure~\ref{fig:ABC_2}.
\end{example}
\begin{example}\label{ex:ABC_3}
If $A = (0,1,0,1,0,1,0)$, $B = (0,1,0,1,1,1,1)$, then $C=(0, 0, 1, 0, 2, 1, 3, 2, 2, 2, 1, 1, 0)$.
A visualization  of $A$, $B$ and $C$ is shown in Figure~\ref{fig:ABC_3}.
\end{example}

Given the definition of non-negative convolution, we want to solve the following sparse non-negative convolution problem.
\begin{definition}
Given vectors $A,B \in \R_+^n$, we want to recover a vector $D \in \mathbb{R}^n$ such that:
\begin{align*}
    \supp(D) = & ~ \supp_{\geq c_1}(A \star B) \\
    D_j = & ~ (A \star B)_j + o(1) \mathrm{~~} \forall j \in \supp_{\geq c_1}(A \star B)
\end{align*}
\end{definition}

We state our main result in the following  two theorems:

Theorem~\ref{thm:approx_sparse_conv_informal} shows that we can compute approximate sparse  non-negative convolution for $A \star B$.
\begin{theorem}[Approximate sparse convolution, informal version of Theorem~\ref{thm:approx_sparse_conv}]\label{thm:approx_sparse_conv_informal}
Let $c_1 = \Omega(1)$ and $c_2 = o(n^{-2})$.  
Suppose that the Assumption~\ref{ass:approx_sparse} holds.

There is an Algorithm~\ref{alg:approx_sparse_conv} that runs in time
\begin{align*}
    O(k \log (n) \log(k)\log(k/\delta)).
\end{align*}
recovers all index in $\supp_{\geq c_1}(A \star B)$ with point-wise error of $o(1)$, i.e. 
\begin{itemize}
    \item $\supp(D) = \supp_{\geq c_1}(A \star B)$,
    \item $D_j = (A \star B)_j + o(1)$ for all $j \in \supp_{\geq c_1}(A \star B)$.
\end{itemize}
holds with probability at least $1-\delta$.
\end{theorem}

Theorem~\ref{thm:iterative_correct_intro} shows that we can iteratively correct the error for the approximate sparse non-negative convolution.
\begin{theorem}[Informal version of Theorem~\ref{thm:iterative_correct}]\label{thm:iterative_correct_intro}
Let $c_1 = \Omega(1)$ and $c_2 = o(n^{-2})$.

Suppose that the Assumption~\ref{ass:approx_sparse} holds.

There is an algorithm (Algorithm~\ref{alg:iterative_correct}) that runs in time 
\begin{align*}
O(k \log(n) \log^2(k) (\log(1/\delta) + \log\log(k)))
\end{align*}
recovers all index in $\supp_{\geq c_1}(A \star B)$ correctly, i.e. 
\begin{itemize}
\item $\supp(D) = \supp_{\geq c_1}(A \star B)$  
\item $D_j = (A \star B)_j $ for all $j \in \supp_{\geq c_1}(A \star B)$
\end{itemize}
holds with probability at least $1-\delta$.
\end{theorem}

%%%%%%%%%%%%%%%%%%%%%%%%%%
\begin{definition}[Cyclic convolution]
The {cyclic convolution} of two length-$n$ vectors $A, B$ is the length-$n$ vector $A \star_n B$ with 
\begin{align*}
	(A \star_n B)_i := \sum_{j=0}^{n-1} A_j B_{(i-j) \mod n}.
\end{align*}
\end{definition}

We define support as follows:
\begin{align*}
    \supp(A) ~ & := \{ i \in [n] : A_i \neq 0 \}
\end{align*}
We define $\ell_0$ norm,
\begin{align*}
    \|A\|_0 ~ & := |\supp(A)|.
\end{align*}
We define $\ell_{\infty}$ norm as follows,
\begin{align*}
    \|A\|_\infty ~ & := \max_{i \in [n]} |A_i|.
\end{align*}

\begin{definition}[Generalized norm and support]
For vector $A \in \R^n$, we define its $\geq C$-norm and $\leq c$-norm as follows:
\begin{itemize}
    \item $\|A\|_{\geq C} := \sum_{i \in [n] } 1 [ A_i \geq C ]$. 
    \item $\|A\|_{\leq c} := \sum_{i \in [n] } 1 [ A_i \leq c]$.
\end{itemize}
We define a matrix $A \in \R^n$ with $\geq C$-support as 
\begin{align*}
    \supp(A)_{\geq C} := \{ i \in [n] : A_i \geq C \}.
\end{align*}
\end{definition}

\begin{definition}[Rounding]
Define function $\mathrm{int}(\cdot): \R \mapsto \mathbb{N}$ which rounds a number to its closest integer.
\end{definition}

\begin{definition}[Affine operator]
We say $\iota$ is an affine operator if 
\begin{align*}
\iota(A) - \iota(B) = \iota(A-B).
\end{align*}
\end{definition}

In the following, when we refer to ``isolated'' or ``non-isolated'' elements, we only consider those in $\supp_{\geq c_1}(A)$.
\begin{definition}[Isolated index]\label{def:isolated_index}
 Let $g(x) = x ~\mathrm{mod}~ p$ where $p$ is a random prime in the range $[m, 2m]$. 
We say that an index $x \in \supp_{\geq c_1}(A \star B)$ is ``isolated'' if there is no other index $x' \in \supp_{\geq c_1}(A \star B)$ with 
\begin{align*}
    g(x') \in (g(x) +\{-2p, -p, 0, p, 2p \}) \mod m.
\end{align*}
\end{definition}

\begin{lemma}[Hash function,\cite{bfn21}]\label{lem:hash_function}
When combined with the ideal hash function $\iota$ gives 
\begin{align*}
\iota(\partial(A \star B))=\iota(\partial A) \star_m \iota(B)+\iota(A) \star_m \iota(\partial B).
\end{align*}

The $b$-th coordinate of this vector is
\begin{align*}
\iota(\partial(A \star B))_b=\sum_{i: \iota(i)=b} i \cdot(A \star B)_i
\end{align*}
which can be accessed by computing the length- $m$ convolutions $\iota(\partial A) \star_m \iota(B)$ and $\iota(A) \star_m \iota(\partial B)$ and adding them together. By setting $m=O(k)$, we can now infer a constant fraction of elements $i \in \operatorname{supp}(A \star B)$ by performing the division
\begin{align*}
\frac{\iota(\partial(A \star B))_b}{\iota((A \star B))_b}=\frac{\sum_{i: \iota(i)=b} i \cdot(A \star B)_i}{\sum_{i: \iota(i)=b}(A \star B)_i}
\end{align*}
for all $b \in[m]$. This yields the locations of all isolated elements in $\operatorname{supp}(A \star B)$ under $\iota$.
%\Guangyi{Is this lemma or definition?}\Guangyi{Do I need to reference it?} \Zhao{You can call it lemma for now. Please refer to it, whenever you call it.}
\end{lemma}

We will leverage the following lemma of hashing modulo a random prime during our analysis.

\begin{lemma}[Hashing Modulo a Random Prime, Lemma 8.2 of \cite{bfn21}] \label{lem:prime-hashing-basics}

With modular hashing, we let the hash function $g(x) = x ~\mathrm{mod}~ p$ where $p$ is a random prime in the range $[m, 2m]$, the value $x$ is an integer hash code generated from the key. 

Then the following properties hold:
\begin{description}
\item[Universality:] For distinct keys $x, y \in [U]$: 
\begin{align*}
    \Pr[g(x) = g(y)] \leq 2 \log(U) / m.
\end{align*}

\item[Affinity:] For arbitrary keys $x, y$:
\begin{align*}
    g(x) + g(y) = g(x + y) \mod p.
\end{align*}

\end{description}
\end{lemma}

We also generalize the definition on $g(x)$ to $g(A)$ for a vector $A\in \R^n$. Let $g(x) = x \mod p$. Then $g(A) \in \R^p$ where 
\begin{align*}
    g(A)_i := \sum_{j\in [n],~g(j)=i} A_j.
\end{align*}

\begin{figure*}[!ht]
\subfloat[$p =2$]{\includegraphics[width=0.30\textwidth]{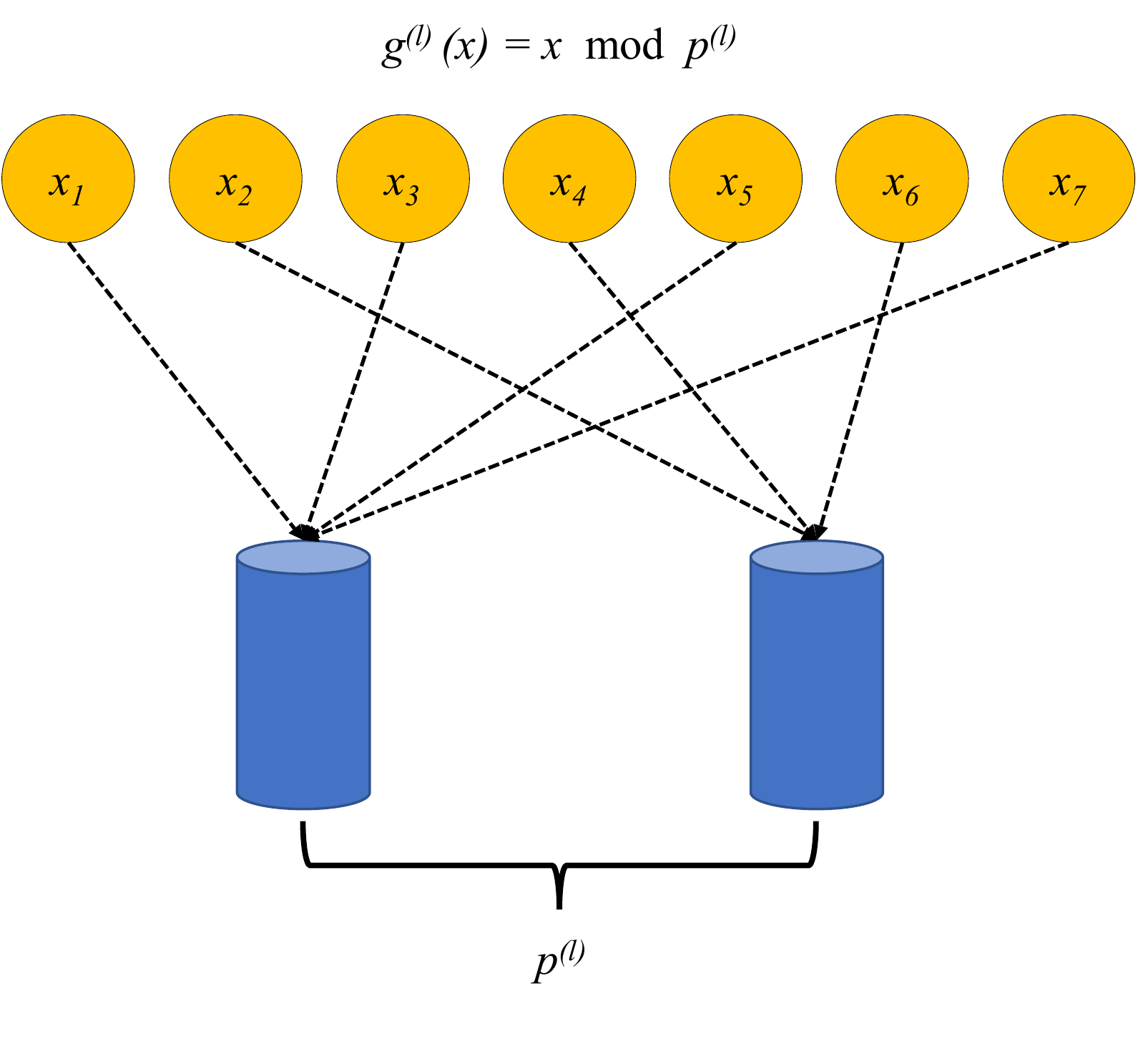}\label{fig:x_mod_p_2}} 
\hspace{1pt}
\subfloat[$p=3$]{\includegraphics[width=0.30\textwidth]{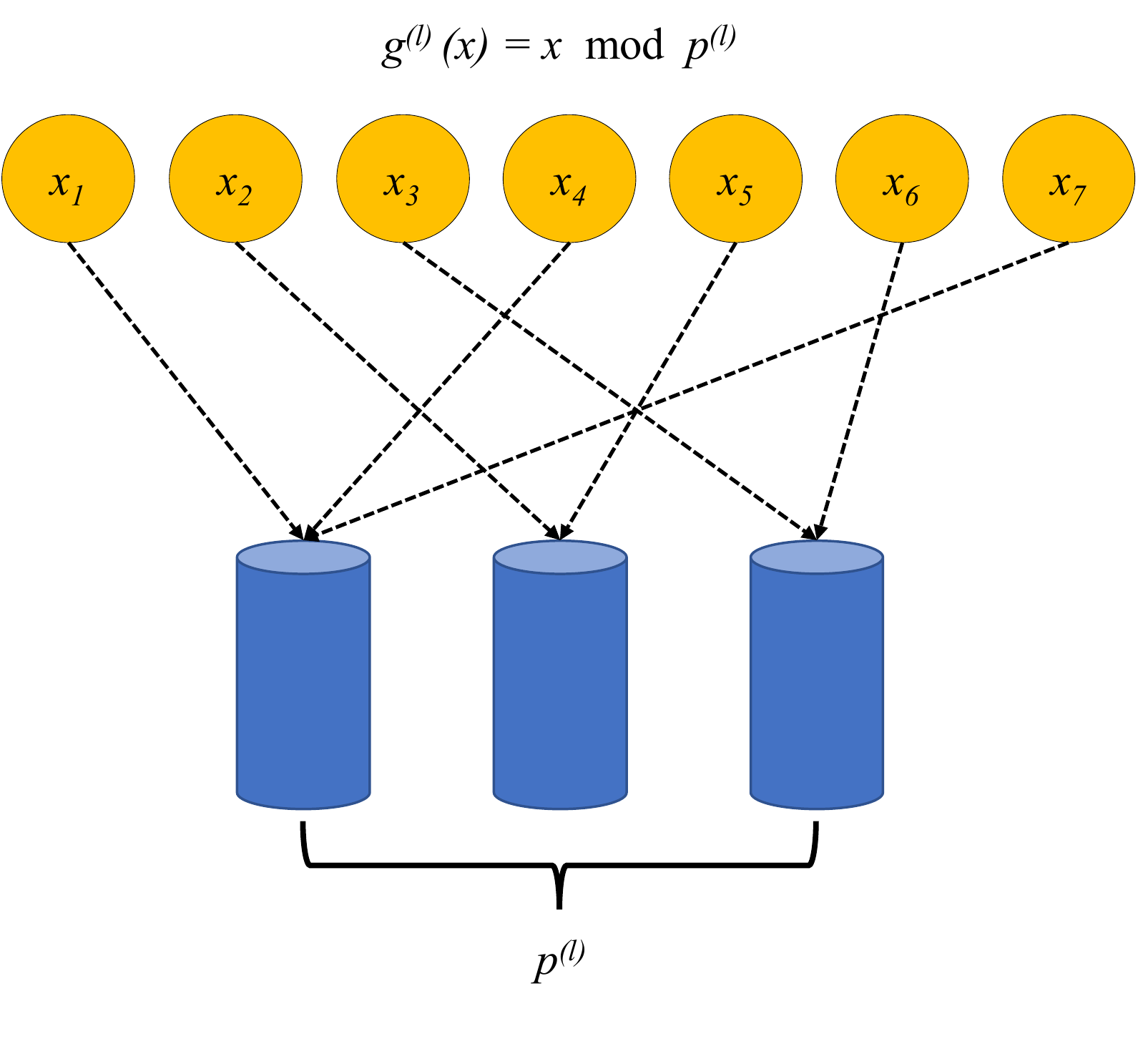}\label{fig:x_mod_p_3}}
\hspace{1pt}
\subfloat[$p=5$]{\includegraphics[width=0.31\textwidth]{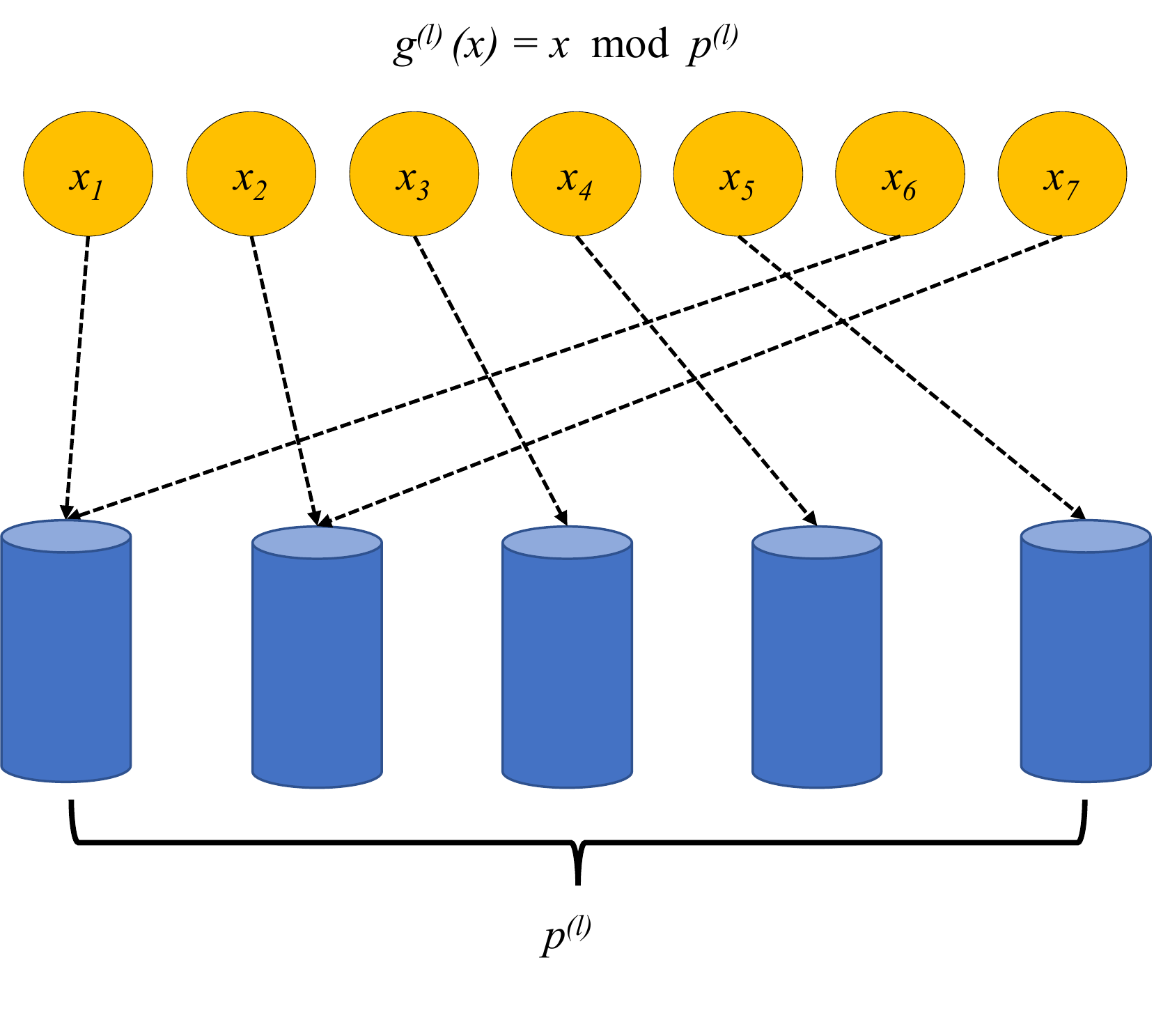}\label{fig:x_mod_p_5}} 
\caption{Visualization of $g(x) = x \mod p$. There are seven elements.}
\end{figure*}

A visualization of $g(x) = x \mod p$  and Algorithm~\ref{alg:approx_sparse_conv} Line~\ref{line:x_mod_p} is shown in Figure~\ref{fig:x_mod_p_3}.

We will also use the Hoeffding bound to obtain high success probability.

\begin{lemma}[Hoeffding bound \cite{h63}]\label{lem:hoeffding}
Let $X_1,\cdots,X_n$ be $n$ independent bounded variables in $[a_i,b_i]$. Let $X:=\sum_{i=1}^n X_i$, then we have
\begin{align*}
    \Pr[|X - \E[X]| \geq t] \leq 2\exp\left(-\frac{2t^2}{\sum_{i=1}^n (b_i-a_i)^2}\right).
\end{align*}

\end{lemma}

%% file: sparse_conv.tex
\section{Approximate sparse non-negative convolution}\label{sec:approx_sparse_conv}

In this section, we present our proposed approximate sparse convolution algorithm architecture as well as corresponding lemmas and proofs.  We first describe our approximate sparse non-negative convolution algorithm in Algorithm~\ref{alg:approx_sparse_conv} and theorem stating the  approximation guarantee and running time complexity.

In the following lemma, we want to prove for all indexes $i \in \supp_{\geq c_1}(A \star B)$, it is isolated in at least $L/2$ hashing functions with high probability.
\begin{lemma}\label{lem:isolated_L_2}
Suppose that the Assumption~\ref{ass:approx_sparse} holds,
with probability at least $1-\delta$, for all $i \in \supp_{\geq c_1}(A \star B)$, $i$ is isolated in at least $L/2$ hashing functions in $\{g^{(l)}\}_{l=1}^L$. 
\label{lem:half_isolated}
\end{lemma}

\begin{proof}
Recall that $A,B \in \R_+^n$ and there exist 
$c_1 = \Omega(1)$ and $c_2 = o(n^{-2})$
such that 
\begin{align*}
\|A \star B\|_{\geq c_1} = k
\end{align*}
and 
\begin{align*}
\|A \star B\|_{\leq c_2} = n-k
\end{align*}

according to Assumption~\ref{ass:approx_sparse}.

Let $g(x) = x ~\mathrm{mod}~ p$ where $p$ is a random prime in the range $[m, 2m]$. 
By Lemma~\ref{lem:prime-hashing-basics}, we have  
\begin{align}\label{eq:i_is_non_isolated}
    \Pr[i \text{ is non-isolated}] \notag 
    \leq & ~ | \supp_{\geq c_1} (A \star B) | \cdot \Pr[ g(x) = g(y) ]  \notag \\
    \leq & ~ k \cdot  \Pr[ g(x) = g(y) ]  \notag \\
    \leq & ~ k \cdot 2 \log(U)  / m \notag \\
    = & ~ \frac{2k \log n}{m} \notag \\
    \leq & ~ \frac{1}{C \log k} \notag \\
    \leq & ~ \frac{1}{4}
\end{align}
 
where the first step follows from definition of isolated index, the second step follows from $| \supp_{\geq c_1} (A \star B) | =k$, the third step follows from $\Pr[g(x) = g(y)] \leq 2 \log(U) / m$, the forth step follows from $U=n$, the fifth step follows from $m \geq C \cdot k ( \log n ) \cdot (\log k)$
, the last step follows from $C \geq 4$ and $\log k \geq 1$.

For every fixed $i \in \supp_{\geq c_1 }(A\star B)$, since hash functions are i.i.d chosen, by Lemma~\ref{lem:hoeffding}, with probability at least $1-\delta/k$, 
\begin{align}\label{eq:lm41_pf2}
    \frac{1}{L}\sum_{l=1}^L \mathbf{1}[i \text{ is non-isolated in }g^{(l)}] 
    \leq & ~ \Pr[i \text{ is non-isolated}] + 2\sqrt{\frac{\log{(k/\delta)}}{L}} \notag \\
    \leq & ~ 1/4 + 2\sqrt{\frac{\log{(k/\delta)}}{L}} \notag \\
    \leq & ~ 1/4 + 1/4 \notag \\
    \leq & ~ 1/2  .
\end{align}
where the first step follows from $\frac{1}{L}\sum_{l=1}^L \mathbf{1}[i \text{ is non-isolated in }g^{(l)}] \leq  \Pr[i \text{ is non-isolated}] + 2\sqrt{\frac{\log{(k/\delta)}}{L}}$, 
where the second step follows from Eq.~\eqref{eq:i_is_non_isolated}, the third step follows from $L \geq 100 \log (k/\delta)$, and the last step follows from simple algebra.

Therefore  it follows by union bound for all $i \in \supp_{\geq c_1 }(A\star B)$.
This completes the proof.
\end{proof}

\begin{algorithm*}[!ht]\caption{Approximate sparse non-negative convolution.}
\label{alg:approx_sparse_conv}
\begin{algorithmic}[1]
\State {\bf data structure}
\State {\bf members}
    \State \hspace{4mm} Vectors $A, B \in \R_+^n$,
    \State \hspace{4mm} and an integer $k$ such that
    Assumption~\ref{ass:approx_sparse} is hold.
    %$\supp(A \star B)_0 \leq k$% and $U = \poly(k)$
\State {\bf end members}
\\
\Procedure{SparseConvolution}{$A, B \in \R_+^n$}
\State  Let {$m \leftarrow {O(k \log(n)}{\log(k)})$} 
\State Let {$L \leftarrow O(\log(k/\delta))$}

\For {$l \leftarrow 1, \dots, L $} \Comment{Do $L$ parallel recovery.}
    \State Randomly pick a prime $p^{(l)} \in [m, 2m]$ and let $g^{(l)}(x) = x ~\mathrm{mod}~ p^{(l)}$ \label{line:x_mod_p}
    \State $V^{(l)} \leftarrow g^{(l)}(A) \star g^{(l)}(B)$ using FFT  \label{line:compute_V} 
     \State $W^{(l)} \leftarrow g^{(l)}(\partial A) \star g^{(l)}(B) + g^{(l)}(A) \star g^{(l)}(\partial B) $ using FFT \label{line:compute_W}
    \For {$i \in \supp_{\geq c_1}(V^{(l)})$} \label{line:loop_cell}
        \State $x \leftarrow W^{(l)}_i / V^{(l)}_i$ \label{line:detect_x}
        \If {$|x - \mathrm{int}(x)| = o(1)$}\Comment{Recover if $x$ is very close to an integer.}
        \State $C.\append(\mathrm{int}(x), V^{(l)}_i)$ \label{line:add_recovery}
        \EndIf
    \EndFor
\EndFor

\State Let $I$ be the set of index $i$ where $i$ occurred as some $(i,*)$ in $C$ \label{line:non_repeating_elements}
\For {$i\in I$} \Comment{Enumerate all possible indexes in $C$}
\State Let $F_i$ be the multi-set of all values $v$ that $(i,v)$ occurs in $C$.  \label{line:compute_F}
\State \Comment{If $(i,v)$ occurs multiple time in $C$, $v$ should appear same amount of time in $F_i$.} 
\State $D_{i} \leftarrow \mathrm{median}(F_i)$ \label{line:compute_median}
\EndFor

\State \Return $D$
\EndProcedure

\State {\bf end data structure}
\end{algorithmic}
\end{algorithm*}

The goal of the following lemma is to prove that the value $V_{i}^{l}$ computed in \textsc{SparseConvolution} in Algorithm~\ref{alg:approx_sparse_conv} satisfies $V^{(l)}_i = (A \star B)_{\mathrm{int}(x)} + o(1)$.

\begin{lemma}\label{lem:sparse_conv_V_approx}
Consider $l\in [L]$ and its corresponding hash function $g^{(l)} : [n] \rightarrow [p^{(l)}]$.
Let 
\begin{align*}
    V^{(l)} & := g^{(l)}(A) \star g^{(l)}(B)\\
    W^{(l)} & := g^{(l)}(\partial A)\star g^{(l)}(B) + g^{(l)}(A) \star g^{(l)}(\partial B)
\end{align*}
as defined in Line~\ref{line:compute_V},~\ref{line:compute_W}.

Let $i\in \supp_{\geq c_1} (A\star B)$ be some coordinate and let 
\begin{align*}
x := W_i^{(l)}/V_i^{(l)}
\end{align*}
as stated in Line~\ref{line:detect_x}.

Suppose $i$ is isolated with respect to $g^{(l)}$,  then  $x$ satisfies 
$|x - \mathrm{int}(x)| = o(1)$
and the $V_i^{(l)}$ in Algorithm~\ref{alg:approx_sparse_conv} Line~\ref{line:compute_V} satisfies 
\begin{align*}
V^{(l)}_i = (A \star B)_{\mathrm{int}(x)} + o(1).
\end{align*}
\end{lemma}

\begin{proof}
There exists unique $\hat x \in \supp_{\geq c_1}(A \star B)$ with $g^{(l)}(\hat x) = i$.
In this case, we have 
\begin{align}\label{eq:lm42_proof}
V^{(l)}_i 
=  & ~ \sum_{y: \iota^{(l)}(y) = i} (A \star B)_{y} \notag \\
= & ~  (A \star B)_{\hat x} + o(1),
\end{align}

where the first step follows from $V^{(l)}_i = \sum_{y: \iota^{(l)}(y) = i} (A \star B)_{y} $ (see Algorithm~\ref{alg:approx_sparse_conv}),
the second step follows from $\sum_{y: \iota^{(l)}(y) = i} (A \star B)_{y} =   (A \star B)_{\hat x} + o(1)$.

Next, we can rewrite $x$ as follows: 
\begin{align*}
        x
        = & ~ \frac{W_i^{(l)}}{ V_i^{(l)} } \\
        = & ~ \frac{\iota^{(l)} (\partial(A \star B))_i}{\iota^{(l)} (A \star B)_i} \\
        = & ~  \frac{\sum_{y: \iota^{(l)}(y) = i} y \cdot (A \star B)_y}{\sum_{y: \iota(x) = i} (A \star B)_y} \\
        = & ~ \hat x + o(1).
\end{align*}
where the first step follows from definition of $x$, the second step follows from definition of $W_i^{(l)}$ and $V_i^{(l)}$, the second step follows from Eq. \eqref{eq:def_partial} and Lemma \ref{lem:hash_function}, the third step follows from Eq. \eqref{eq:lm42_proof}.

Therefore 
\begin{align*}
|x - \hat x| = o(1)
\end{align*}
and 
\begin{align*}
|V^{(l)}_i -(A \star B)_{\hat x}| = o(1).
\end{align*}
This completes the proof.
\end{proof}

With the premise of Lemma~\ref{lem:isolated_L_2} and Lemma~\ref{lem:sparse_conv_V_approx}, it is possible to prove the approximation guarantee and time complexity of \textsc{SparseConvolution} in Algorithm~\ref{alg:approx_sparse_conv} in Theorem~\ref{thm:approx_sparse_conv}.

\begin{theorem}[Approximate sparse convolution, formal version of Theorem~\ref{thm:approx_sparse_conv_informal}]\label{thm:approx_sparse_conv}
Suppose that the Assumption~\ref{ass:approx_sparse} holds, let $c_1 = \Omega(1) \text{ and } c_2 = o(n^{-2})$.
Algorithm~\ref{alg:approx_sparse_conv} runs in time
\begin{align*}
    O(k \log (n) \log(k)\log(k/\delta))
\end{align*}
recovers all index in $\supp_{\geq c_1}(A \star B)$ with point-wise error of $o(1)$, i.e.
\begin{itemize}
    \item $\supp(D) = \supp_{\geq c_1}(A \star B) $
    \item $D_j = (A \star B)_j + o(1)$
for all $j \in \supp_{\geq c_1}(A \star B)$
\end{itemize}
holds with probability at least $1-\delta$.

\end{theorem}

\begin{proof}
We initiate the high probability event in Lemma~\ref{lem:half_isolated}. 
In this event, for each $i \in I$, we have 
$L/2 \leq |F_i| \leq L$,
so we must have $D_{i} = \mathrm{median}(F_i) = V^{(l)}_j$
for some $l$ and $j$ such that $i$ is isolated and $i = W_j/V_j$.  Then by Lemma~\ref{lem:sparse_conv_V_approx}, this implies 
$D_{i} = (A \star B)_{i}+o(1)$
and $\supp_{\geq c_1}(A \star B) \subseteq I$.
Hence the first statement is proven. 

For running time, we notice the followings
\begin{itemize}
    \item Line~\ref{line:compute_V}-\ref{line:compute_W} takes $O(k \log(n) \log^2(k))$ time. 
    \item Line~\ref{line:loop_cell}-\ref{line:add_recovery} takes one pass of all recovery, which takes $O(mL)$ time. 
    \item Line~\ref{line:non_repeating_elements} and Line~\ref{line:compute_F} takes totally $O(|C|)=O(mL)$ time.
    We could build up a perfect hash function, scanning though all pairs in $C$ and take out all possible indexes and its corresponding value in linear time.
    \item Line~\ref{line:compute_median} takes $O(mL)$ time in total.For each $i\in I$, take median value of a set $F_i$ takes $O(|F_i|)$ time. 
    
    Because $|F_i| \leq L$ and $|I|\leq m$, overall it takes 
    \begin{align*}
        O(\sum_{i \in I} |F_i|) = O(mL)
    \end{align*}
    time.  
\end{itemize}

Since $m = {O(k \log(n)}{\log(k)})$, $L = O(\log(k/\delta))$, 
then
\begin{align*}
    mL = O(k \log(n) \log(k) \log(k/\delta)).
\end{align*}

To sum up, the total running time is:
\begin{align*}
    O(k\log (n) \log^2(k) + mL) = O(k \log (n) \log(k)\log(k/\delta)).
\end{align*}
\end{proof}

%% file: iterative_correct_error.tex
\section{Iteratively correcting the errors}\label{sec:approx_sparse_conv_iter_correcting_error}

The previous section illustrates the architecture of algorithms to approximate convolution computation. In this section, we will show how to iteratively corrects the errors in Algorithm~\ref{alg:iterative_correct}.

\begin{theorem}[Formal version of Theorem~\ref{thm:iterative_correct_intro}]\label{thm:iterative_correct}
Let $c_1 = \Omega(1)$ and $c_2 = o(n^{-2})$.
Suppose Assumption~\ref{ass:approx_sparse} holds.

There is an algorithm 
(Algorithm~\ref{alg:iterative_correct}) that runs in time
\begin{align*}
O(k \log(n) \log^2(k) (\log(1/\delta) + \log\log(k)))
\end{align*}

recovers all index in $\supp_{\geq c_1}(A \star B)$ correctly, i.e. 
\begin{itemize}
\item   $\supp(D) = \supp_{\geq c_1}(A \star B)$ 
\item $D_j = (A \star B)_j $ for all $j \in \supp_{\geq c_1}(A \star B)$
\end{itemize}
holds with probability at least $1-\delta$.
\end{theorem}

\begin{proof}
The proof comes from Lemma~\ref{lem:correct_iterative} and Lemma~\ref{lem:time_iterative}. This completes the proof.
\end{proof}

\begin{algorithm*}[!ht]\caption{Sparse non-negative convolution.}
\label{alg:iterative_correct}
\begin{algorithmic}[1]
\State {\bf data structure}
\State {\bf members}
\State \hspace{4mm} Vectors $A, B \in \R_+^n$
\State \hspace{4mm} Integer $k$ such that $\supp(A \star B)_0 \leq k$% and $U = \poly(k)$
\State {\bf end members}
\\
\Procedure{SparseConvolution}{$A, B \in \R_+^n$}
\State  {$m \leftarrow {8k \log(n)}{\log^2(k)}$}
\State $L \gets \Theta(\log \log k)$
\For {$l \leftarrow 1, \dots, L$}
    \State $R_l \gets {2\log(2L/\delta) / 1.5^{l-1}}$
    \For {$r \leftarrow 1,\dots, R_l$} \label{line:inner-boosting-start}
        \State Randomly pick a prime $p \in [m, 2m]$ 
        \State $g_r(x) := x ~\mathrm{mod}~ p$
        \State $V_r \leftarrow g_r(A) \star g_r(B) - g_r(C^{l - 1})$ using FFT \label{line:compute_V_iterative} 
        \State $W_r \leftarrow g_r(\partial A) \star g_r(B) + g_r(A) \star g_r(\partial B) - g_r(\partial C^{l-1})$ using FFT \label{line:inner-boosting-end}
    \EndFor
    \State $r^* \gets \arg\max_{r \in [R_l]} |\supp_{\geq c_1} (V_r)|$
    \State $g \gets g_{r^*}$, $V \gets V_{r^*}$, $W \gets W_{r^*}$ \label{line:keep}
    \State $C^l \leftarrow C^{l-1}$
    \For {$i \in \supp_{\geq c_1}(V)$} \label{line:loop_cell_start}
        \State $x \leftarrow W_i / V_i$ \label{line:detect-x}
        \If {$|x - \mathrm{int}(x)| = o(1)$}
            \State $C^l_x \leftarrow C^l_x + V_i$ \label{line:update-Cell} 
            % \Comment{$\mathrm{int}(\cdot)$ rounds a real number to its closest integer.}
        \EndIf
    \EndFor\label{line:loop_cell_end}
\EndFor
\State $C \gets C^{L}$
\State\Return $C $
\EndProcedure

\State {\bf end data structure}
\end{algorithmic}
\end{algorithm*}

We use the following Lemma from \cite{bfn21}. 
Since this result only depends on hash function and Lines~\ref{line:inner-boosting-start}--\ref{line:inner-boosting-end}, the proof is similar.
\begin{lemma}[Lemma 8.3 in \cite{bfn21}, Most Indices are Isolated] \label{lem:isolation-most-error-corr}
Let $\ell$ be any level. If 
\begin{align*}
\|A \star B - C^{l-1}\|_{\geq c_1} \leq \frac{2^{-1.5^{l-1}} k}{\log^2(k)},
\end{align*}
then with probability $1 - \delta / (2L)$, there will be at most
\begin{align*}
    \frac{2^{-1.5^l} k}{2\log^2(k)}
\end{align*}
non-isolated elements at level $l$.
\end{lemma}

\begin{definition}[$L$ and $R_l$]
We define $L$ as 
\begin{align*}
 L := \Theta( \log \log k )
\end{align*}
For each $l \in [L]$, we define
\begin{align*}
R_l := \Theta(\log(L/\delta)) / 1.5^{l-1}.
\end{align*}
\end{definition}

\begin{definition}[Residual and derivative of residual]\label{def:iterative_residual}
Consider $r\in [R_l]$
and its corresponding hash function $g_r : [n] \rightarrow [p_r]$. 
We define $V_r$ and $W_r$ as follows
\begin{align*}
    V_r & := g_r(A) \star g_r(B)- g_r(C^{l-1})\\
    W_r & := g_r(\partial A) \star g_r(B) + g_r(A) \star g_r(\partial B) - g_r(\partial C^{l-1}).
\end{align*}
Also see in Line~\ref{line:compute_V_iterative} and~\ref{line:inner-boosting-end} in Algorithm~\ref{alg:iterative_correct}. 
\end{definition}

\begin{definition}[Finding the largest residual]
We define 
\begin{align*}
    r^* := \arg\max_{r \in [R_l]} |\supp_{\geq c_1} (V_r)|.
\end{align*}
\end{definition}

\begin{lemma}[Isolated Indices are Recovered] \label{lem:isolation-recovery-error-corr}

Denoting the number of non-isolated elements at level~$l$ by~$r$, we have 
\begin{align*}
\|A \star B - C^l\|_{\geq c_1} \leq 2r.
\end{align*}
\end{lemma}

\begin{proof}
Focus on arbitrary $l$, and assume that we already picked a hash function $g$ in Algorithm~\ref{alg:iterative_correct} Lines~\ref{line:inner-boosting-start}--\ref{line:keep}. 
In Definition~\ref{def:iterative_residual}, we have provided the definition of $V$ and $W$.
By the affinity of $g$ it holds that 
$V = g(A \star B - C^{l-1})$,
by additionally using the product rule
$W = g(\partial(A \star B - C^{l-1}))$.

Now focus on an arbitrary $i \in [n]$. There are three cases:

{\bf Case 1.} 

For all $x \in [n]$ with $g(x) = i$, there is 
\begin{align*}
0 \leq (A \star B - C^{l-1})_x \leq c_2 \cdot n^{l-1}.
\end{align*}
    
In this case, we have 
\begin{align*} 
V_i = & ~ \sum_{x: \iota(x) = i} (A \star B)_x \\
\leq & ~ n \cdot c_2 \\
= & ~  o(1).
\end{align*}
Thus $i \notin \supp_{\geq c_1}(V)$.

{\bf Case 2.} 

There exists unique $x$, such that 
\begin{align*}
x \in \supp_{\geq c_1}(A \star B - C^{l-1})
\end{align*}
with $g(x) = i$.
    
    In this case, we have 
    \begin{align*}
      V_i = \sum_{x: \iota(x) = i} (A \star B)_x 
      = x + o(1)  
    \end{align*}
    and
    \begin{align*}
        \frac{\iota (\partial(A \star B))_i}{\iota (A \star B)_i} 
        = & ~ \frac{\sum_{x: \iota(x) = i} x \cdot (A \star B)_x}{\sum_{x: \iota(x) = i} (A \star B)_x} \\
        = & ~ x + o(1).
    \end{align*}
    Therefore we successfully recover 
    \begin{align*}
    C_x^l = \mathrm{int}( V_1) =  (A \star B)_x.
    \end{align*}
    
{\bf Case 3.} 

There exists multiple $x \in \supp_{\geq c_1}(A \star B - C^{l-1})$ with $g(x) = i$.
In this case, we have
    \begin{align*}
        V_i = \sum_{x: \iota(x) = i} (A \star B - C^{l-1})_x
    \end{align*}
    and 
    \begin{align*}
        \frac{\iota (\partial(A \star B - C^{l-1}))_i}{\iota (A \star B - C^{l-1})_i}
        =  \frac{\sum_{x: \iota(x) = i} x \cdot (A \star B - C^{l-1})_x}{\sum_{x: \iota(x) = i} (A \star B - C^{l-1})_x}.
    \end{align*}
    If $V_i \geq c_1$ and $\dfrac{\iota (\partial(A \star B - C^{l-1}))_i}{\iota (A \star B - C^{l-1})_i} = \hat x$, then 
\begin{align*}
    (A \star B - C^{l})_{\hat x} = \Omega(1).
\end{align*}
    
Otherwise this iteration does not recover any coordinate in $A \star B - C^{l-1}$.

\end{proof}

\begin{lemma}[Correctness of Algorithm~\ref{alg:iterative_correct}]\label{lem:correct_iterative}
In Algorithm~\ref{alg:iterative_correct}, we can correctly outputs $C = A \star B$ with probability $1 - \delta$.
\end{lemma}
\begin{proof}
We show that with probability $1-\delta$ it holds that
 \begin{align*}
 \|A \star B-C^{\ell}\|_{\geq c_1} \leq 2^{-1.5^{\ell}} k \log ^{-2}(k) 
\end{align*}
for all levels $\ell$.

At the last level, 
\begin{align*}
    L= \log _{1.5} \log k =  O(\log \log k),
\end{align*}
we must have 
\begin{align*}
\|A \star B-C^{\ell}\|_{\geq c_1}=0 .
\end{align*}
and thus 
\begin{align*}
A \star B=C^{\ell}=C.
\end{align*}

The proof is by induction on $\ell \in[L+1]$. 

For $\ell=0$, the statement is true assuming that the \textsc{SparseConvolution} in Algorithm~\ref{alg:iterative_correct} with parameter $\delta / 2 \leq \log ^{-2}(k) / 2$ succeeds. 

For $\ell>1$, we appeal to the previous lemmas: By the induction hypothesis we assume that
\begin{align*}
    \| A \star B- C^{\ell-1} \|_{\geq c_1} \leq 2^{-1.5^{\ell-1}} k \log ^{-2}(k).
\end{align*}

Hence, by Lemma~\ref{lem:isolation-most-error-corr}, the algorithm picks a hash function $g$ under which only $2^{-1.5^{\ell}} k \log ^{-2}(k) / 2$ elements are non-isolated at level $\ell$.

By Lemma~\ref{lem:isolation-recovery-error-corr} it follows that 
\begin{align*}
    \|A \star B-C^{\ell}\|_{\geq c_1} \leq 2^{-1.5^{\ell}} k \log ^{-2}(k),
\end{align*}
 which is exactly what we intended to show.

 For $\ell=0$, the error probability is $\delta / 2$. For any other level, the error probability is $1-\delta /(2 L)$ by Lemma~\ref{lem:isolation-most-error-corr} and there are $L$ such levels in total. 
 
 Taking a union bound over these levels, we can obtain the desired error probability of $1-\delta$. This completes the proof.
\end{proof}

\begin{lemma}[Time complexity of Algorithm~\ref{alg:iterative_correct}]\label{lem:time_iterative}
There is an algorithm (Algorithm~\ref{alg:iterative_correct}) that can compute $C = A \star B$
in 
\begin{align*}
O(k \log(n) \log^2(k) (\log(1/\delta) + \log\log(k)))
\end{align*}
time.
\end{lemma}

\begin{proof}

The running time of \textsc{SparseConvolution} in Algorithm~\ref{alg:iterative_correct} can be computed in the following steps:
\begin{itemize}
    \item Line~\ref{line:compute_V_iterative} and Line~\ref{line:inner-boosting-end} takes $O(k \log(n) \log^2(k))$ time to do the FFT. 
    
    We can bound the number of iterations by :
    \begin{align*}
        \sum_{\ell=1}^{L}[\frac{2 \log (2 L / \delta)}{1.5^{\ell-1}}]
        = & ~ (2 \log (2 L / \delta)) \sum_{\ell=1}^{L} \frac{1}{1.5^{\ell-1}} \\
        \leq & ~ (2 \log (2 L / \delta))\cdot 10\\
        = & ~ O(\log (L / \delta)) ,
    \end{align*}
where the second step follows from the sum of infinite geometric series.

Therefore, the total time spent on FFT is 
\begin{align*}
    O(k \log(n) \log^2(k) \log(1/\delta)).
\end{align*}

    \item Line~\ref{line:loop_cell_start} to Line~\ref{line:loop_cell_end} take 
    \begin{align*}
        O(mL) = O(k \log(n) \log^2(k) \log\log(k)).
    \end{align*}
    
\end{itemize}

Therefore, the overall time complexity is:
\begin{align*}
    ~ & O(k \log(n) \log^2(k) \log(1/\delta))
   + O(k \log(n) \log^2(k) \log\log(k))\\
    = & ~ O(k \log(n) \log^2(k) (\log(1/\delta) + \log\log(k)))
\end{align*}
This completes the proof.

\end{proof}

%% file: conclusion.tex
\section{Conclusion}\label{sec:conclusion}

The computation of the convolution $A \star B$ of two vectors of dimension $n$ is considered to be one of the most important and fundamental computational primitives in a wide variety of fields. Utilizing the Fast Fourier Transform, which has a time complexity of $O(n \log n)$, is the traditional way to solve the problem of non-negative convolution might have a very sparse representation, which is a property that we can use to our advantage to speed up the computation to obtain the result. 

In this paper, we show that for approximately $k$-sparse case, we can approximately recover the all index in $\supp_{\geq c_1}(A \star B)$ with point-wise error of $o(1)$ in $O(k \log (n) \log(k)\log(k/\delta))$ time. We further show that we can iteratively correct the error and recover all index in $\supp_{\geq c_1}(A \star B)$ correctly in 
$O(k \log(n) \log^2(k) (\log(1/\delta) + \log\log(k)))$
time.